\documentclass[12pt,twoside,chap]{paper}
\usepackage{axodraw}
\usepackage{minitoc,cite,fancyhdr}
\usepackage{bbm} 
\usepackage{euscript,amssymb,latexsym} 
\usepackage{epsf,epsfig,graphics,subfigure,graphicx}
\usepackage{amsmath}
      
\usepackage{eepic}
       
\makeatletter
\def\fmslash{\@ifnextchar[{\fmsl@sh}{\fmsl@sh[0mu]}}
\def\fmsl@sh[#1]#2{%
  \mathchoice
    {\@fmsl@sh\displaystyle{#1}{#2}}%
    {\@fmsl@sh\textstyle{#1}{#2}}%
    {\@fmsl@sh\scriptstyle{#1}{#2}}%
    {\@fmsl@sh\scriptscriptstyle{#1}{#2}}}
\def\@fmsl@sh#1#2#3{\m@th\ooalign{$\hfil#1\mkern#2/\hfil$\crcr$#1#3$}}
\makeatother
\global\arraycolsep=2pt 

\makeatletter
\def\cleardoublepage{\clearpage\if@twoside \ifodd\c@page\else%
\hbox{}%
\thispagestyle{empty}%
\newpage%
\if@twocolumn\hbox{newpage}\fi\fi\fi}
\makeatother

\begin{document}
\title{Next-to-Leading-Order Monte Carlo Simulation of 
       Diphoton Production in Hadronic Collisions}
\author{\bf{\normalsize{Luca D'Errico}}
\\ 
\vspace{-0.3 cm} \it{\normalsize{Institut f\"ur Theoretische Physik,}}
\\ 
\vspace{-0.1 cm} \it{\normalsize{University of Karlsruhe, KIT, 76128, Germany;}}
\\
\vspace{-0.3 cm} \it{\normalsize{Institute of Particle Physics Phenomenology, Department of Physics,}}
\\ 
\vspace{-0.1 cm} \it{\normalsize{University of Durham, DH1 3LE, UK;}}
\\
\it{\normalsize{Email:} }
\sf{\normalsize{derrico@particle.uni-karlsruhe.de}}
\\
\\
\vspace{-0.1 cm} \bf{\normalsize{Peter Richardson}} 
\\ 
\vspace{-0.3 cm} \it{\normalsize{Institute of Particle Physics Phenomenology, Department of Physics,}}
\\
\vspace{-0.1 cm} \it{\normalsize{University of Durham, DH1 3LE, UK;}}
\\
\it{\normalsize{Email:} }
\sf{\normalsize{peter.richardson@durham.ac.uk}}
}
\maketitle
\vspace{-14cm}
\begin{flushright}
KA-TP-11-2011\\
SFB/CPP-11-30\\
MCNET-11-15\\
DCPT/11/68\\
IPPP/11/34 
\end{flushright}
\vspace{12.5cm}
\begin{abstract}
We present a method, based on the positive weight next-to-leading-order matching formalism~(POWHEG), to simulate
photon production processes at next-to-leading-order~(NLO). This technique is applied to the simulation
of diphoton production in hadron-hadron collisions. The algorithm consistently combines the parton shower
and NLO calculation, producing only positive weight events. The simulation includes both the photon fragmentation contribution
and a full implementation of the truncated shower required to correctly describe soft emissions in an angular-ordered
parton shower.
\end{abstract}
\thispagestyle{empty}
\newpage
\thispagestyle{plain}
\setcounter{page}{1}
---------------------------------------------------------------------------------------------------------
\tableofcontents
---------------------------------------------------------------------------------------------------------
\section{Introduction}

The production of photons via perturbative processes is very important for both the search for the
Higgs boson and other new physics, via photon pair production, and for the study of QCD and
experimental effects, in particular the jet-energy scale,
in the production of a photon in association with a jet. To
study these processes in detail in hadron-hadron collisions we need an accurate Monte Carlo simulation.
In this paper we present a new approach for the simulation of
these processes and illustrate it with the simulation of photon pair production.

Monte Carlo event generators simulate events by combining fixed-order matrix elements,
parton showers and hadronization models. The first programs used leading-order~(LO) matrix elements,
together with the parton shower approximation which describes soft and collinear emission. 
Recently different approaches correcting the emission of high transverse momentum, $p_{T}$, partons
have been introduced\footnote{See Ref.\,\cite{Buckley:2011ms} for a recent review of the older
techniques~\cite{Sjostrand:1986hx, Bengtsson:1987rw, Norrbin:2000uu, Miu:1998ju,Corcella:2000bw,Corcella:2002jc,Seymour:1991xa,Seymour:1994ti,Corcella:1998rs,Corcella:1999gs,Seymour:1994df,Seymour:1994we,Gieseke:2003hm, Gieseke:2004af, Hamilton:2006ms, Gieseke:2006ga, Bahr:2008tx, Gieseke:2011na}
and techniques for improving the simulation of multiple hard QCD
radiation~\cite{Catani:2001cc, Krauss:2002up, Lonnblad:2001iq, Schalicke:2005nv, Krauss:2005re, Lavesson:2005xu, Mrenna:2003if, Mangano:2002ea, Alwall:2007fs,Hoeche:2009rj,Hamilton:2009ne}.}.

Different algorithms have been  developed to provide a better description of the hardest emission,
including the full next-to-leading-order cross section. In the approach of Frixione and Webber~(MC@NLO) \cite{Frixione:2002ik,Frixione:2010wd}, the parton shower approximation is subtracted
from the real emission contribution to the next-to-leading-order cross section
and combined with the virtual correction. This method was successfully applied to
many different processes~\cite{Frixione:2005vw,Frixione:2007zp,Frixione:2008yi,LatundeDada:2007jg, LatundeDada:2009rr,Papaefstathiou:2009sr,Torrielli:2010aw,Frixione:2010ra}. 
However, this approach has two drawbacks: it generates weights that are not positive definite and
its implementation depends on the parton shower algorithm.

These drawbacks have been addressed with a new method introduced by Nason~\cite{Nason:2004rx, Frixione:2007vw},
the POsitive Weight Hardest Emission Generator~(POWHEG) approach. This method generates positive weights
and is implemented in a way that does not depend on the details of the parton shower algorithm.
Nevertheless, the parton shower algorithm must have a well defined structure: 
a \textit{truncated shower} simulating wide angle soft emission;
followed by the emission with highest transverse momentum~($p_{T}$);
followed by a \textit{vetoed shower} simulating softer radiation. 
The hardest emission is generated by a Sudakov form factor that includes the 
full matrix element for real emission. The truncated shower generates emission at a
higher scale~(in the evolution variable of the parton shower), while the vetoed shower simulates radiation
at a lower evolution scale than the one at which the hardest emission is generated.
The POWHEG method has been successfully applied to a wide range of
processes~\cite{Nason:2006hfa,Frixione:2007nw,LatundeDada:2006gx,Alioli:2008gx,Hamilton:2008pd,Alioli:2008tz,Hamilton:2009za,Alioli:2009je,Hoche:2010pf,Alioli:2010xd,Nason:2009ai,Re:2010jg,Re:2010bp,Alioli:2010qp,Alioli:2010xa,Oleari:2010nx,Hamilton:2010mb,Oleari:2011ey,Kardos:2011qa,Melia:2011gk}\footnote{
There has also been some work combining either many NLO matrix elements~\cite{Lavesson:2008ah} or the NLO matrix elements with subsequent emissions matched to leading-order matrix elements~\cite{Hamilton:2010wh,Hoche:2010kg} with the parton shower.}. 

These approaches have yet to be applied to processes involving the production of photons due to
the complications which arise in the experimental measurement, simulation and calculation
at higher orders in perturbation theory of these processes. Collider experiments do not 
measure \textit{inclusive} photons because of the high background due to the production of photons in meson decays.
Indeed, the inclusive production rate of high $p_{T}$ $\pi^{0}$, $\eta$, $\omega$ mesons is orders of magnitude bigger than
for direct photon production. For this reason the experimental selection of direct photons requires
the use of an isolation cut.  Different criteria for the isolation of photons include:
the cone approach \cite{Baer:1990ra, Aurenche:1989gv}, the democratic approach \cite{Glover:1993xc} and
the smooth isolation procedure \cite{Frixione:1998jh}.
In fixed-order calculations this contribution is included using the measured photon fragmentation function, the
probability of a parton fragmenting to produce a photon with a given fraction of the parent parton's momentum, whereas
Monte Carlo simulations instead rely on the parton shower and hadronization models to simulate this contribution.
This presents a problem in simulating these processes at NLO where some of the singularities in the
real emission processes are absorbed into photon fragmentation function in fixed-order calculations.
In this paper we will present a method for simulating these processes using the POWHEG approach which
still relies on the parton shower and hadronization models to simulate the photon fragmentation contribution.
This approach is similar in its philosophy to the method of Ref.\,\cite{Hoeche:2009xc} for combining
leading-order matrix elements and the parton shower.

We illustrate this approach with the simulation of diphoton ($\gamma\gamma$) production.
Diphoton production is important as it provides a large background for the discovery of the Higgs boson decaying
into a pair of photons, for both the Tevatron~\cite{Abazov:2009kq}
and LHC experiments~\cite{Aad:2009wy, Ball:2007zza}. It is also an important background in new physics
models, for example in heavy resonance models~\cite{Mrenna:2000qh}, models with extra spatial dimensions~\cite{Han:1998sg}
and cascade decays of heavy new particles \cite{Giudice:1998bp}. 
Experimental measurements of $\gamma\gamma$ production have a long history in fixed-target \cite{Bonvin:1990kx,Bonvin:1988yu,Begel:1999py} and 
collider experiments \cite{Albajar:1988im,Alitti:1992hn,Abe:1992cy,Acosta:2004sn,Abazov:2010ah}.

The theoretical understanding of diphoton production and precise measurements of the differential
cross section are therefore not only important for the discovery of new phenomena but also as a check of the
validity of the predictions of perturbative quantum chromodynamics (pQCD)
and soft-gluon resummation methods. 

The dominant production method for direct photon pairs is leading order $q\bar{q}$ scattering~($q\bar{q}\to\gamma\gamma$),
although the formally next-to-next-to-leading-order, $\mathcal{O}(\alpha_{s}^{2})$, gluon-gluon fusion ($gg\to\gamma\gamma$)
process via a quark-loop diagram~\cite{Berger:1983yi}
can be important, and even comparable to the leading-order contribution at low diphoton mass~($M_{\gamma\gamma}$) \cite{Abazov:2010ah},   
due to the large gluon parton distribution function.

The $\mathcal{O}(\alpha_{s})$ corrections to the $q\bar{q}\to\gamma\gamma$ process 
includes the \mbox{$q\bar{q}\to\gamma\gamma g$},\linebreak \mbox{$gq\to\gamma\gamma q$} and \mbox{$g\bar{q}\to\gamma\gamma\bar{q}$} processes
and corresponding virtual corrections. Moreover, the contribution where the final parton is collinear to a photon
is calculated in terms of the quark and gluon fragmentation function into a photon~\cite{Berger:1983yi, LlewellynSmith:1978dc}.
Given the behaviour of the latter functions, $\sim\frac{\alpha}{\alpha_{s}}$, these terms contribute to the same order
as $q\bar{q}\to\gamma\gamma$. The QCD corrections to the process are well known in the 
literature~\cite{Baer:1990ra, Aurenche:1987fs, Gordon:1994ut, Aurenche:1985yk, Bailey:1992br,DelDuca:2003uz}. 
Fixed-order Monte Carlo programs, such as \textsf{JETPHOX} \cite{Catani:2002ny} and \textsf{DIPHOX} \cite{Binoth:1999qq}, provide simulation
for direct photon production together with the implementation of isolation cuts.

The present paper is organized as follows. In Sect.~\ref{sec1} we introduce the POWHEG formulae useful
for the description of our approach and our treatment of the photon fragmentation contribution.
The calculation of the leading-order kinematics with NLO accuracy in the POWHEG approach is
discussed in Sect.~\ref{sec2}. In Sect.~\ref{sec3} we describe the procedure used to generate the hardest emission.
We show our results in Sect.~\ref{sec4} and finally present our conclusions in Sect.~\ref{sec5}.

\section{The POWHEG method} 
\label{sec1}

In the POWHEG approach the NLO differential cross section for a given N-body process is
\begin{equation}
\mathrm{d}\sigma=\bar{B}(\Phi_{B})\mathrm{d}\Phi_{B}
\left[\Delta_{R}(0)+\frac{R(\Phi_{B},\Phi_{R})}{B(\Phi_{B})}
\Delta_{R}(k_{T}(\Phi_{B},\Phi_{R}))\mathrm{d}\Phi_{R}\right]\text{,}\label{lo_scs}
\end{equation}
where $\bar{B}(\Phi_{B})$ is defined as
\begin{equation}
\bar{B}(\Phi_{B})=B(\Phi_{B})+V(\Phi_{B})
+\int \left[ R(\Phi_{B},\Phi_{R})-C(\Phi_{B},\Phi_{R})\right]
\mathrm{d}\Phi_{R}\text{,} \label{bbar1}
\end{equation}
$B(\Phi_{B})$ is the leading-order contribution, $\Phi_{B}$ the N-body phase-space
variables of the LO Born process whereas  $\Phi_{R}$ are the
radiative variables describing the phase space for the emission of an extra parton.
The real contribution, $R(\Phi_{B},\Phi_{R})$,
 is the matrix element including the radiation of an
additional parton multiplied by the relevant parton flux factors,
and is regulated by subtracting the counter terms $C(\Phi_{B},\Phi_{R})$
which contain the same singularities as $R(\Phi_{B},\Phi_{R})$.
In practice the counter term is usually composed of a sum over
a number of terms, $D^i(\Phi_{B},\Phi_{R})$, each of which regulates one of the singularities in the matrix element
using approaches of either Catani and Seymour~(CS) \cite{Catani:1996vz} or
Frixione, Kunszt and Signer~(FKS) \cite{Frixione:1995ms}, \textit{i.e.}\ \mbox{$C(\Phi_{B},\Phi_{R})=\sum_{i}D^{i}(\Phi_{B},\Phi_{R})$}.
The finite contribution $V(\Phi_{B})$ includes the
virtual loop corrections and the counter terms integrated over the real emission
variables, which cancel the singularities from the loop corrections, and the collinear remnant from
absorbing the initial-state singularities into the parton distribution functions.

The modified Sudakov form factor is defined in terms of the real emission matrix element
\begin{equation}
\Delta_{R}(p_{T})=\exp \left [-\int \mathrm{d}\Phi_{R}
\frac{R(\Phi_{B},\Phi_{R})}{B(\Phi_{B})}
\theta(k_{T}(\Phi_{B},\Phi_{R})-p_{T}) \right]\text{,} \label{mod_sud}
\end{equation}
where $k_{T}(\Phi_{B},\Phi_{R})$ is equal to the transverse momentum of the
emitted parton in the soft and collinear limits.

The POWHEG method is based on two steps: the N-body configuration is generated according to $\bar{B}(\Phi_{B})$
and then the hardest emission is generated using the Sudakov form factor given in Eqn.~\ref{mod_sud}.
Since $\bar{B}(\Phi_{B})$ is defined as the NLO differential cross section integrated over the radiative variables,
the event weight will not be negative.

If the parton shower algorithm is ordered in transverse momentum we would generate the hardest emission
first and then evolve the $N+1$ parton final-state system using the shower forbidding any emission with
transverse momentum higher than that of the hardest emission. 
On the contrary for shower simulations which are ordered in other variables, such as angular ordering in \textsf{Herwig++}~\cite{Bahr:2008pv,Gieseke:2011na}
,
the hardest emission is not necessarily the first one. For this reason the shower must be split into a
truncated shower describing soft emission at higher evolution scales, 
the highest $p_T$ emission and vetoed showers simulating emissions at lower evolution scales;
however, constraints are imposed to guarantee that the transverse momentum of the emitted particles is smaller
than the one corresponding to the hardest emission \cite{Nason:2004rx,Frixione:2007vw}.

In order to use this procedure for processes involving photons where the real emission matrix elements
contain both QCD singularities from the emission of soft and collinear gluons and QED singularities
from the radiation of soft and collinear photons we need to make some modifications to the approach.
We start by writing the real emission piece as
\begin{equation}
R(\Phi_{B},\Phi_{R})=R_{\rm QED}(\Phi_{B},\Phi_{R})+R_{\rm QCD}(\Phi_{B},\Phi_{R})\text{,}
\end{equation}  
where 
\begin{subequations}
\begin{equation}
R_{\rm QED}(\Phi_{B},\Phi_{R})=\frac{\sum_{i}D^{i}_{\rm{QED}}}{\sum_{j}D^{j}_{\rm{QED}}+\sum_{j}D_{\rm{QCD}}^{j}}R(\Phi_{B},\Phi_{R})
\end{equation} 
contains the collinear photon emission singularities and 
\begin{equation}
R_{\rm{QCD}}(\Phi_{B},\Phi_{R})=\frac{\sum_{i}D^{i}_{\rm{QCD}}}{\sum_{j}D^{j}_{\rm{QED}}+\sum_{j}D_{\rm{QCD}}^{j}}R(\Phi_{B},\Phi_{R})
\end{equation}
\end{subequations}
contains the singularities associated with $\rm{QCD}$ radiation.\footnote{In practice the counter terms can be negative in some regions
and we choose to use their magnitude in this separation in order to ensure that the real contributions are positive.}
Here the counter terms have been split into those $D^i_{\rm QCD}$ which regulate the singularities from QCD radiation and those
$D^i_{\rm QED}$ which regulate the singularities due to photon radiation.

We can regard the real QCD emission terms as part of the QCD corrections to the leading-order process, whereas the
QED contributions are part of the photon fragmentation contribution coming from a leading-order process with one
less photon and an extra parton. We therefore modify the next-to-leading-order cross section for processes with photon production
giving 
\begin{eqnarray}
\rm{d}\sigma&=&\left\{B(\Phi_{B})+V(\Phi_{B})+\int\left[ R_{\rm{QCD}}(\Phi_{B},\Phi_{R})-\sum_{i}D^{i}_{\rm{QCD}}(\Phi_{B},\Phi_{R})\right]
\mathrm{d}\Phi_{R}\right\}\rm{d}\Phi_{B}\notag \\
&&+R_{\rm{QED}}(\Phi_{B},\Phi_{R})\mathrm{d}\Phi_{R}\rm{d}\Phi_{B}\text{.} 
\end{eqnarray}
There should also be an additional non-perturbative contribution with the convolution of the photon fragmentation
function and the leading-order process with one less photon and an extra parton.

We can now write the cross section for photon production processes in the POWHEG approach in 
the same way as in Eqn.\,\ref{lo_scs}
\begin{eqnarray}
\mathrm{d}\sigma&=&\phantom{+}\!
\bar{B}(\Phi_{B})\mathrm{d}\Phi_{B}
\left[\Delta_{\rm QCD}(0)+\frac{R_{\rm QCD}(\Phi_{B},\Phi_{R})}{B(\Phi_{B})}
\Delta_{\rm QCD}(k_{T}(\Phi_{B},\Phi_{R}))\mathrm{d}\Phi_{R}\right]\\
&& +B'({\Phi'}_B)\mathrm{d}{\Phi'}_B
\left[\Delta_{\rm QED}(0)+\frac{R_{\rm QED}({\Phi'}_B,{\Phi'}_R)}{B'({\Phi'}_B)}
\Delta_{\rm QED}(k_{T}({\Phi'}_B,{\Phi'}_R))\mathrm{d}{\Phi'}_R\right]
\text{,}\nonumber
\end{eqnarray}
where $\bar{B}(\Phi_{B})$ is now defined as
\begin{equation}
\bar{B}(\Phi_{B})=\left\{B(\Phi_{B})+V(\Phi_{B})+\int\left[ R_{\rm{QCD}}(\Phi_{B},\Phi_{R})-\sum_{i}D^{i}_{\rm{QCD}}(\Phi_{B},\Phi_{R})\right]
\mathrm{d}\Phi_{R}\right\}\rm{d}\Phi_{B}
\label{photon_Bbar}
\end{equation}
and $B'({\Phi'}_B)$ is the leading-order contribution for the process with an extra parton and one less photon with
${\Phi'}_B$ and ${\Phi'}_R$ being the corresponding Born and real emission phase-space variables.

The Sudakov form factor for QCD radiation is
\begin{subequations}
\begin{equation}
\Delta_{\rm QCD}(p_{T})=\exp \left [-\int \mathrm{d}\Phi_{R}
\frac{R_{\rm QCD}(\Phi_{B},\Phi_{R})}{B(\Phi_{B})}
\theta(k_{T}(\Phi_{B},\Phi_{R})-p_{T}) \right]\text{,}
\label{eqn:qcd_sud}
\end{equation}
and the Sudakov form factor for QED radiation is
\begin{equation}
\Delta_{\rm QED}(p_{T})=\exp \left [-\int \mathrm{d}{\Phi'}_R
\frac{R_{\rm QED}({\Phi'}_B,{\Phi'}_R)}{B'({\Phi'}_B)}
\theta(k_{T}({\Phi'}_B,{\Phi'}_R)-p_{T}) \right]\text{.}
\label{eqn:qed_sud}
\end{equation}
\end{subequations}

Both the direct photon production and the non-perturbative fragmentation contribution are
correctly included. The non-perturbative fragmentation contribution
is simulated by the parton shower from the $B'({\Phi}'_B)$ contribution when there is no
hard QED radiation.

The POWHEG algorithm is implemented for photon production processes using the following
procedure.
\begin{itemize}
\item First select either a direct photon production or a fragmentation event
using $\bar{B}(\Phi_B)$ and $B'({\Phi'}_B)$ and the competition method to correctly generate
the relative contributions of the two different processes.
\item For a direct photon production process:
\begin{itemize}
\item generate the hardest emission using the Sudakov form factor in Eqn.\,\ref{eqn:qcd_sud};
\item directly hadronize non-radiative events;
\item map the radiative variables parameterizing the emission into the evolution
 scale, momentum fraction and azimuthal angle, $(\tilde{q}_{h},z_{h},\phi_{h})$,
 from which the parton shower would reconstruct identical momenta;
\item generate the $N$-body configuration from $\bar{B}(\Phi_{B})$
 and evolve the radiating parton from the starting scale
 down to $\tilde{q}_{h}$ using the truncated shower;
\item insert a branching with parameters $(\tilde{q}_{h},z_{h},\phi_{h})$ into the
 shower when the evolution scale reaches $\tilde{q}_{h}$;
\item generate $p_{T}$ vetoed showers from all the external legs. 
\end{itemize}
\item For a fragmentation contribution:
\begin{itemize}
\item generate the hardest QED emission using the Sudakov form factor in Eqn.\,\ref{eqn:qed_sud};
\item directly shower and hadronize non-radiative events, forbidding any perturbative QED radiation
in the parton shower generating the\linebreak non-perturbative fragmentation contribution;
\item for events with QED radiation map the radiative variables parameterizing the emission into the evolution
 scale, momentum fraction and azimuthal angle, $(\tilde{q}_{h},z_{h},\phi_{h})$,
 from which the parton shower would reconstruct identical momenta;
\item generate the $N$-body configuration from $B'({\Phi'}_B)$
 and evolve the radiating parton from the starting scale
 down to $\tilde{q}_{h}$ using the truncated shower, but allowing QCD radiation with $p_T$ greater
than that of the hardest QED emission;
\item insert a branching with parameters $(\tilde{q}_{h},z_{h},\phi_{h})$ into the
 shower when the evolution scale reaches $\tilde{q}_{h}$;
\item generate the shower from all external legs forbidding QED radiation, but not QCD radiation,
above the $p_T$ of the hardest emission. 
\end{itemize}
\end{itemize}
This procedure now includes the QCD corrections to the leading-order direct photon
production process and both the perturbative QED corrections to the photon fragmentation
contribution and the non-perturbative contribution are simulated by the parton shower.

In the next two sections we will describe how we implement this approach in
\textsf{Herwig++} for photon pair production.

\begin{figure}[t]
\centering
\includegraphics[width=0.5\textwidth]{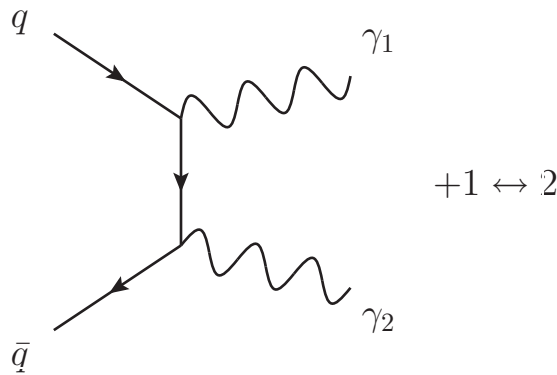}
\caption{Diphoton production at leading-order.}\label{LO_pp}
\end{figure}
%
%
\section{Calculation of \boldmath{$\bar{B}(\Phi_{B})$}}
\label{sec2}
\begin{figure}[t]
\centering
\includegraphics[width=\textwidth]{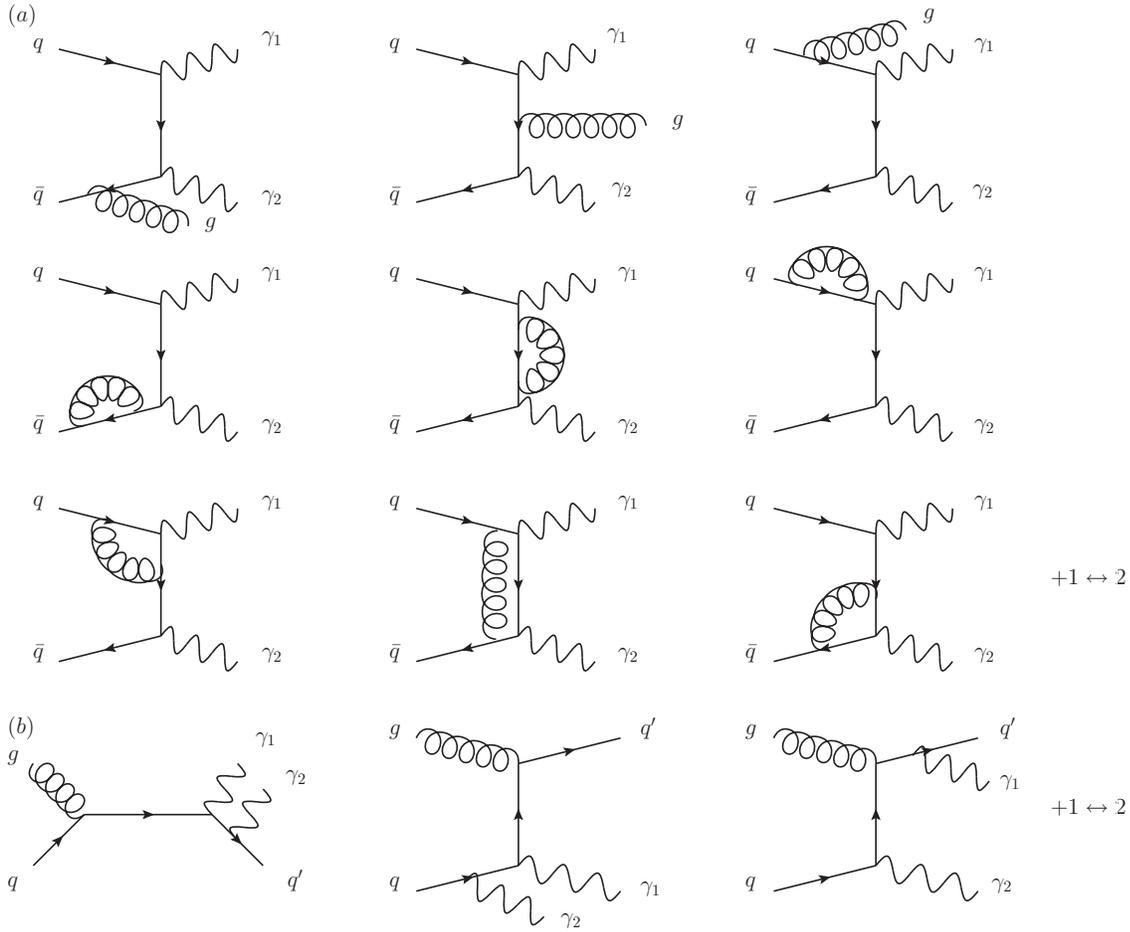}
\caption{Diphoton production at next-to-leading-order. In (a) the real and virtual Feynman diagrams contributing to the $q\bar{q}\to\gamma\gamma$ subprocess are shown while in (b) the real diagrams for $gq$ initiated process are given.}\label{NLO_pp}
\end{figure}

\begin{figure}[t]
\centering
\includegraphics[width=\textwidth]{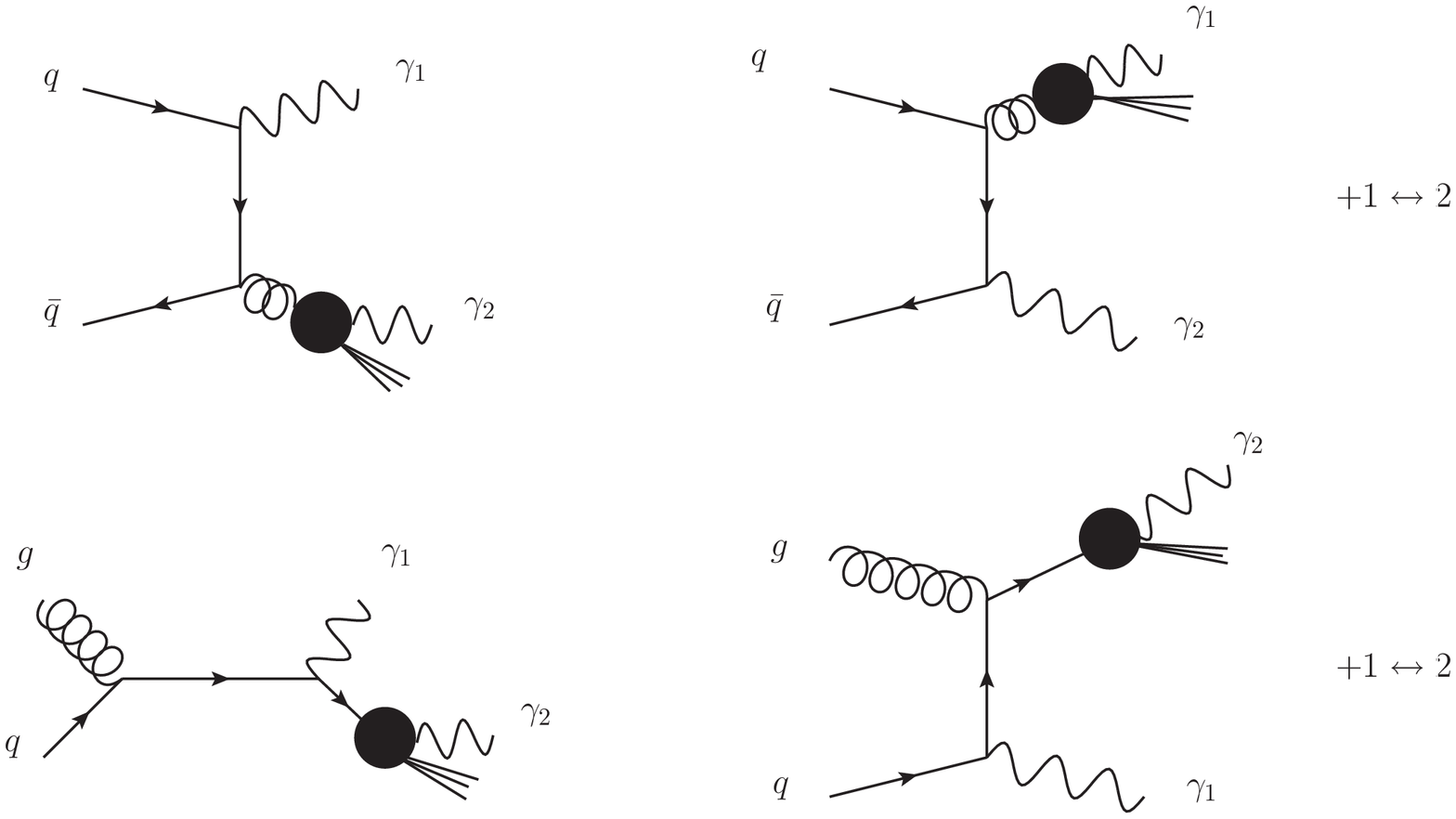}
\caption{Bremsstrahlung contribution for diphoton production.}\label{NLO_pponefrag} 
\end{figure}

In this section we describe the $\mathcal{O}(\alpha_s)$ corrections to diphoton production.
At leading-order, $\gamma\gamma$-production is described by the Feynman diagram illustrated in Fig.\,\ref{LO_pp}.
Next-to-leading-order contributions yield $\mathcal{O}(\alpha_{s})$ corrections coming from $q\bar{q}\to\gamma\gamma g$, $gq\to\gamma\gamma q$ and $g\bar{q}\to\gamma\gamma\bar{q}$, together with the corresponding virtual corrections, as shown in Fig.\,\ref{NLO_pp}.  
These subprocesses contain QED singularities, corresponding to configurations where the final-state
parton becomes collinear to a photon, which  do not cancel when summing up the real and the virtual pieces of the cross section.
As described in the previous section they are formally absorbed into a quark ($G_{\gamma q}(z,\mu^{2})$) or gluon ($G_{\gamma g}(z,\mu^{2})$) fragmentation function into photons, which define the probability of finding a photon carrying longitudinal momentum fraction $z$ in a quark or gluon jet at scale $\mu$ for a given factorization scheme.
This QED singular component is called the \textit{Bremsstrahlung} or single fragmentation contribution.
In our approach it is treated separately and simulated by showering the $gq\to\gamma q$ or $g\bar{q}\to\gamma \bar q$ within the
Monte Carlo algorithm, see Fig.\,\ref{NLO_pponefrag}, as described in the previous section.
At next-to-leading-order the same configuration appears in any subprocess in which a quark (gluon) undergoes a cascade
of successive collinear splittings ending up with a quark-photon (gluon-photon) splitting. These singularities are factorized
to all orders in $\alpha_{s}$, according to the factorization theorem. When the fragmentation scale $\mu$ is chosen higher
than any other hadronic scale, \textit{i.e.} $\mu \sim$ 1 GeV, these functions behave roughly
as $\frac{\alpha}{\alpha_{s}(\mu^{2})}$ and therefore they contribute at leading-order.  

For a full study at NLO accuracy, the $\mathcal{O}(\alpha_{s})$ corrections to the Bremsstrahlung contribution need to be calculated. Moreover, these corrections in their turn yield the leading-order contribution of the double fragmentation type process; in the latter case, both photons result from the collinear fragmentation of a parton. However, these corrections are out of the scope of the present work and are not considered here.

\subsection{Real emission contribution}
\label{realemissionsect}

In order to calculate the real emission contribution to $\bar{B}(\Phi_B)$ we need to specify both
the radiative phase space, $\Phi_R$, and the subtraction counter terms. We choose to use the
dipole subtraction algorithm of Catani and Seymour~\cite{Catani:1996vz} to specify the counter terms
and the associated definition of the real emission phase space as follows.

In the centre-of-mass frame the incoming hadronic momenta are, $P_{\oplus}$ and $P_{\ominus}$, respectively for the
hadrons traveling in the positive and negative $z$-directions. Similarly the momenta of the incoming partons in the Born
process are $\bar p_{\oplus}=\bar x_{\oplus}P_{\oplus}$ and $\bar p_{\ominus}=\bar x_{\ominus}P_{\ominus}$, respectively.
The momenta of the photons in the Born process are $\bar k_{1,2}$ respectively.
The corresponding momenta in the real emission process are $p_{\oplus}$ and $p_{\ominus}$ for the incoming partons
and $k_{1,2,3}$ for the outgoing particles which are chosen such that $k_{1,2}$
are the momenta of the photons and $k_3$ that of the radiated final-state parton.

In the CS approach the real phase space depends on which parton is the emitter of the radiation and which the associated
spectator defining the dipole~\cite{Catani:1996vz}. 
When the parton with momentum $\bar p_{\oplus}$ is the emitter and
that with momenta $\bar p_{\ominus}$ the spectator the full phase space is~\cite{Catani:1996vz}
\begin{equation} \label{radps} 
{\rm d}\Phi_3={\rm d}\Phi_B{\rm d}\Phi_R = {\rm d}\Phi_B\frac{(k_1+k_2)^2}{16 \pi^2} \, \frac{d \phi_{\oplus}}{2 \pi}
  \,d v_{\oplus} \,\frac{d x}{x} \,\theta (v_{\oplus})\, \theta \!\left( 1 - \frac{v_{\oplus}}{1 - x}
  \right) \theta (x (1 - x)) \, \theta (x - \bar{x}_{\oplus})\text{,}
\end{equation}
where the radiative phase space variables are
\begin{eqnarray}
x=1-\frac{(p_{\oplus}+p_{\ominus})\cdot k_{3}}{p_{\oplus}\cdot p_{\ominus}},\hspace{1 cm}v_{\oplus}=\frac{p_{\oplus}\cdot k_{3}}{p_{\oplus}\cdot p_{\ominus}},\hspace{1 cm}\phi_{\oplus} \text{,}\label{radvar}
\end{eqnarray}
$\phi_{\oplus}$ is the azimuthal angle of the emitted particle around the $\hat{\oplus}$-direction and 
\begin{equation}
x\in [x_{\oplus},1],\hspace{2 cm} v_{\oplus}\in [0,1-x]\text{.}
\end{equation}

In terms of these variables 
\begin{subequations}\label{eqbar}
\begin{eqnarray} 
  p_{\oplus}  = \bar{p}_{\oplus}/x \text{,} \qquad\quad
  p_{\ominus} = \bar{p}_{\ominus} \text{,}\\
  x_{\oplus}  = \bar{x}_{\oplus}/x\text{,} \qquad\quad
  x_{\ominus} = \bar{x}_{\ominus} \text{.}
\end{eqnarray}
\end{subequations} 

It is useful to specify the momentum of the radiated parton in terms of its transverse
momentum, $p_T$, and rapidity, $y$, such that
\begin{equation}
k_3 = p_T\left(\cosh y; \cos\phi_{\oplus}, \sin\phi_{\oplus}, \sinh y\right).
\end{equation} 

Using the definition of $x$ and $v_{\oplus}$ we have 
\begin{equation}\label{eqptdef}
  k_{3} = v_{\oplus} p_{\ominus} + (1 - x - v_{\oplus}) p_{\oplus} + q_{\perp}\text{,}
\end{equation}
where $q_{\perp}$ is the component of the 4-momenta transverse to the beam direction.
The on-shell condition, $k_{3}^{2}=0$, gives
\begin{equation} 
\label{eqptcon}
  -q_{\perp}^2=p_{T}^{2} = 2 p_{\oplus} \cdot p_{\ominus} (1 - x - v_{\oplus}) v_{\oplus}\text{.}
\end{equation}
From Eqn.~\ref{eqptdef} and the definition of rapidity
\begin{equation}
y=\frac{1}{2}\mathrm{ln}\left [\frac{k_{3}^{E}+k_{3}^{z}}{k_{3}^{E}-k_{3}^{z}}\right ]=\frac{1}{2}\mathrm{ln}\left [\frac{(1-x-v_{\oplus})x_{\oplus}}{v_{\oplus}xx_{\ominus}}\right ],\label{rapk3}
\end{equation}
the CS variables are
\begin{equation}
\left\{ 
\begin{array}{c} \label{xandv}
v_{\oplus}=\frac{1}{x_{\ominus}\sqrt{s}}p_{T}e^{-y},\vspace{0.5 cm} \\  
x=\frac{1-\frac{p_{T}}{x_{\ominus}\sqrt{s}}e^{-y}}{1+\frac{p_{T}}{x_{\oplus}\sqrt{s}}e^{y}}.
\end{array}
\right.  
\end{equation} 

This is sufficient to calculate the momentum of the radiated parton, however,
rather than implementing the real emission variables
in the Sudakov form factor in this way and then imposing the $\theta(k_{T}(\Phi_{B},\Phi_{R})-p_{T})$
function it is easier to consider the real emission in terms of the transverse momentum, rapidity
and azimuthal angle of the emitted parton.

The Jacobian for this transformation is 
\begin{equation}
\left | \frac{\partial (x,v)}{\partial p_{T}\partial y}\right | 
=\frac{\frac{2p_{T}}{sx_{\oplus}x_{\ominus}}
\left(1-\frac{p_{T}}
{\sqrt{s}x_{\ominus}}e^{-y}\right)}
{\left(1+\frac{p_{T}}{\sqrt{s}x_{\oplus}}e^{y} \right)^2} = \frac{2p_Tx^2}{sx_{\oplus}x_{\ominus}(1-v_{\oplus})}\text{.}
\end{equation}

The momenta of the photons in the real emission process can then be calculated 
from the Born momenta using
\begin{equation}
  k^{\mu}_r = \Lambda^{\mu}_{\phantom{\mu}\nu}  \bar{k}^{\nu}_r \hspace{2em} \hspace{2em} r = 1, 2,
\end{equation}
where the Lorentz transformation is
\begin{equation}
\Lambda^{\mu}_{\phantom{\mu}\nu} =  g^{\mu}_{\phantom{\mu}\nu} - \frac{2 (K + \bar{K})^{\mu}
  (K + \bar{K})_{\nu}}{(K + \bar{K})^2} + \frac{2 K^{\mu}\bar{K}_{\nu}}{K^2}\, ,
\end{equation}
with
\begin{subequations}\label{bmatrix}
\begin{eqnarray} 
  K &=& p_{\oplus} + p_{\ominus} - k_{3} = k_{1} + k_{2}\text{,}\\ 
  \bar{K} &=& \bar{p}_{\oplus} + \bar{p}_{\ominus}\text{.} 
\end{eqnarray}
\end{subequations}   
The condition $K^{2}=\bar{K}^{2}$ is compatible with the definition of $x$ given in Eqn.~\ref{radvar}.
The kinematic variables for the $\hat{\ominus}$ collinear direction are calculated in a
similar way and they provide a radiative phase space as in Eqn.~\ref{radps}. Moreover,
given the $x_{\oplus}\leftrightarrow x_{\ominus}$ asymmetry of the rapidity in Eqn.~\ref{rapk3},
it is $\left[y\right]_{\ominus}=-\left[y\right]_{\oplus}$. In the rest of the paper we refer to the
collinear direction as $\hat{\rm{O}}=\{\hat{\ominus},\hat{\oplus}\}$, when both components need to be included.

In addition to the real emission variables we need the dipole subtraction terms of Ref.~\cite{Catani:1996vz}. 
In the following $B(\Phi_{B})$ and $B'({\Phi'}_{B})$ are computed in terms of the reduced momenta defined in terms of 
the momenta for the real emission process in Ref.~\cite{Catani:1996vz}.
The QCD singularities from $q\bar{q}\to\gamma\gamma g$ are absorbed by the dipoles
\begin{subequations}
\begin{eqnarray}\label{qeddip1}
D^{qg,\bar{q}}&\equiv& D_{\rm QCD}^{qg}=8\pi{C}_{F}\alpha_{s}(\mu_{R})\frac{1}{2\bar{p}_{\rm \oplus}k_3}\left\{\frac{2}{1-x}-(1+x) \right\}B(\Phi_{B})\text{,}\\
D^{\bar{q}g,q}&\equiv& D_{\rm QCD}^{\bar{q}g}=8\pi{C}_{F}\alpha_{s}(\mu_{R})\frac{1}{2\bar{p}_{\rm \ominus}k_3}\left\{\frac{2}{1-x}-(1+x) \right\}B(\Phi_{B})\text{,} 
\end{eqnarray}
\end{subequations}
where the dipoles $D^{ij,k}$ denote the emitter $i$, emitted parton $j$ and spectator $k$.

The $gq\to \gamma\gamma q$ subprocess involves the QCD dipoles 
\begin{equation}
D^{gq,q}\equiv D_{\rm QCD}^{gq}=8\pi T_{F}\alpha_{s}(\mu_{R})\frac{1}{2\bar{p}_{\oplus}k_3}\left\{1-2x(1-x) \right\}B(\Phi_{B})\text{.} 
\end{equation}

In order to separate the QCD and QED emission we also need the QED dipoles
\begin{subequations}
\begin{eqnarray}
D^{q}_{q\gamma}\equiv D_{\rm QED}^{q\gamma F}&=&8\pi\alpha e_q^2\frac{1}{2k_2k_3\xi}\left\{\frac{2}{1-\xi+z}-2+z \right\}B'({\Phi'}_B)\text{,}\\
D^{q\gamma}_{q}\equiv D_{\rm QED}^{q\gamma I}&=&8\pi\alpha e_q^2\frac{1}{2p_{\ominus}k_3\xi}\left\{\frac{2}{1-\xi+z}-(1+x)\right\}B'({\Phi'}_B)\text{,}
\end{eqnarray}
\end{subequations}
where
\begin{subequations}
\begin{eqnarray}
\xi&=&1-\frac{k_2k_3}{(k_2+k_3)p_{\oplus}}\text{,}\\
z&=&\frac{p_{\oplus}k_2}{(k_2+k_3)p_{\oplus}}\text{,}
\end{eqnarray}
\end{subequations}
and $e_q$ is the charge of the quark $q$ in units of the electron charge. In this case, the radiative phase
space is ${\rm d}\Phi_R^{\prime}(\xi,z,\phi^{\prime})$.
Similar dipoles are included for the $g\bar{q}\to \gamma\gamma \bar{q}$ subprocess. 
We do not include perturbative QED radiation from the $q\bar q \to \gamma g $ subprocess as 
it does not give a perturbative correction to $G_{\gamma g}(z,\mu^{2})$.

In practice we generate the real emission piece as a contribution from each of the incoming partons as
\begin{eqnarray}
\lefteqn{\int\left[ R_{\rm{QCD}}(\Phi_{B},\Phi_{R})-\sum_{i}D^{i}_{\rm{QCD}}(\Phi_{B},\Phi_{R})\right]{\rm d}\Phi^i_R 
= } &\\
&&\sum_{i=\oplus,\ominus} \int\left[\frac{|D^{i}_{\rm{QCD}}|}{\sum_{j}|D^{j}_{\rm{QED}}|+\sum_{j}|D_{\rm{QCD}}^{j}|}R(\Phi_{B},\Phi^i_{R})   -  D^{i}_{\rm{QCD}}(\Phi_{B},\Phi_{R})\right]{\rm d}\Phi^i_R.\nonumber
\end{eqnarray}

For the later generation of the Sudakov form factor it is useful to express the dipoles as
\begin{eqnarray}
D_{\rm QCD}^{I}\equiv\frac{\mathcal{C}_{I}\alpha_{s}(\mu_{R})}{2\pi}\mathcal{D}^{I}B(\Phi_{B})\text{,}
\end{eqnarray} 
where $I=\{qg;\bar{q}g; gq;g\bar{q}\}$,
\begin{subequations}
\begin{eqnarray}
\mathcal{C}_{qg}&=&\mathcal{C}_{\bar{q}g}=C_{F}\text{,}\\
\mathcal{C}_{gq}&=&\mathcal{C}_{g\bar{q}}=T_{F}\text{,}
\end{eqnarray}
\end{subequations}
and
\begin{eqnarray}
D_{\rm QED}^{J}\equiv \frac{\alpha}{2\pi}e_q^2\mathcal{D}_{J}B(\Phi_{B}^{\prime})\text{,}
\end{eqnarray} 
where $J=\{q\gamma F,q\gamma I,\bar{q}\gamma F,\bar{q}\gamma I\}$.
 
\subsection{Virtual contribution and collinear remainders}

The finite piece of the virtual correction is
\begin{equation}
\mathrm{d}\sigma_{V}=\frac{C_{F}\alpha_{s}(\mu_{R})}{2\pi}V(w)B(\Phi_{B})\text{.}
\end{equation}
where the finite contribution of $\mathbf{I}(\epsilon)$~\cite{Catani:1996vz} and the
virtual correction~\cite{Aurenche:1985yk} is
\begin{equation} 
V(w)=\left(3+\mathrm{ln}^2w+\mathrm{ln}^2(1-w)+3\mathrm{ln}(1-w)\right) +\frac{F(w)}{\left(\frac{1-w}{w}+\frac{w}{1-w}\right)}\text{,}
\end{equation}
where $e_{q}$ is the electric charge of quark $q$, and
\begin{eqnarray} 
F(w)&=&2\mathrm{ln}w+2\mathrm{ln}(1-w)+\frac{3(1-w)}{w}(\mathrm{ln}w-\mathrm{ln}(1-w))\notag \\
&+&\left(2+\frac{w}{1-w}\right)\mathrm{ln}^2w+\left(2+\frac{1-w}{w}\right)\mathrm{ln}^2(1-w)\text{,}
\end{eqnarray}
with $w=1+\frac{\hat{t}}{\hat{s}}$, where $\hat s$ and $\hat t$ are the usual Mandelstam variables.

The collinear remainders are
\begin{equation}
\mathrm{d}\sigma_{\rm coll}=\frac{C_{F}\alpha_{s}(\mu_{R})}{2\pi}\frac{f^{m}(x_{\rm{O}},\mu_{F})}{f(x_{\rm{O}},\mu_{F})}  B(\Phi_{B})\text{,}
\end{equation}
where the modified PDF is\footnote{We write the modified PDF for the quark $q$, but a similar expression is valid for an incoming antiquark $\bar{q}$.}
\begin{eqnarray}
f^{m}_{q}(x_{\rm{O}},\mu_{F})&=&\int_{x_{\rm{O}}}^{1}\frac{\mathrm{d}x}{x}\left\{f_{g}\left(\frac{x_{\rm{O}}}{x},\mu_{F}\right)A(x)\right.\notag\\
&+&\left.\left[f_{q}\left(\frac{x_{\rm{O}}}{x},\mu_{F}\right)-xf_{q}(x_{\rm{O}},\mu_{F})\right]B(x)\right. \notag\\
&+&\left.f_{q}\left(\frac{x_{\rm{O}}}{x},\mu_{F}\right)C(x)\right\}
+f_{q}(x_{\rm{O}},\mu_{F})D(x_{\rm{O}})\text{,}\ \ \
\end{eqnarray}
$f_{q}$ and $f_g$ are the quark and gluon PDFs respectively, and
\begin{eqnarray}
A(x)&=&\frac{T_{F}}{C_{F}}\left[2x(1-x)+(x^2+(1-x)^2)\mathrm{ln}\frac{Q^{2}(1-x)^2}{\mu_{F}^{2}x}\right]\text{,}\\
B(x)&=&\left[\frac{2}{1-x}\mathrm{ln}\frac{Q^{2}(1-x)^2}{\mu_{F}^{2}}\right]\text{,}\\
C(x)&=&\left[1-x-\frac{2}{1-x}\mathrm{ln}x-(1+x)\mathrm{ln}\frac{Q^{2}(1-x)^2}{\mu_{F}^{2}x}\right]\text{,}\\
D(x_{\rm{O}})&=&\left[\frac{3}{2}\mathrm{ln}\left(\frac{Q^2}{\mu_{F}^{2}}\right)+2\mathrm{ln}(1-x_{\rm O})\mathrm{ln}\left(\frac{Q^2}{\mu_{F}^2}\right)+2\mathrm{ln}^2(1-x_{\rm{O}})+\frac{\pi^2}{3}-5\right]\text{.} 
\end{eqnarray}
The combined contribution of the finite virtual term and collinear remnants is 
\begin{equation}
\mathrm{d}\sigma_{V+{\rm coll}}=\frac{C_{F}\alpha_{s}(\mu_{R})}{2\pi}\mathcal{V}(\Phi_B)B(\Phi_{B})\text{,}
\end{equation}
where
\begin{equation}
\mathcal{V}(\Phi_{B}) \equiv V(w)+\tilde{V}(x_{\rm{O}},\mu_{F})\text{,}
\end{equation} 
with $\tilde{V}(x_{\rm O},\mu_F)=\frac{f^m(x_{\rm O},\mu_F)}{f(x_{\rm O},\mu_F)}$.
\subsection{Generation of the hard process}\label{subsec3}

The next-to-leading-order simulation of photon pair production in \textsf{Herwig++}
uses the standard \textsf{Herwig++} machinery to generate photon pair and photon plus jet production
in competition. The $\bar{B}$ function is implemented as a reweighting
of the leading-order matrix element as follows:
\begin{enumerate}
\item the radiative variables $\Phi_R\left\{x,v,\phi \right \}$ and $\Phi_R^{\prime}\left\{\xi,z,\phi^{\prime} \right \}$
are transformed into a new set such that the radiative phase space is a unit volume;
\item using the standard \textsf{Herwig++} leading-order matrix element generator, we generate a
leading-order configuration and provide the Born variables $\Phi_{B}$ with an associated weight $B(\Phi_{B})$;
\item the radiative variables $\Phi_{R}$ are generated and $\bar{B}(\Phi_{B})$ sampled
in terms of the unit cube $(\tilde{x},\tilde{v},\tilde{\phi})$, using the Auto-Compensating Divide-and-Conquer~(ACDC) phase-space generator \cite{Lonnblad:2006pt};
\item the leading-order configuration is accepted with a probability proportional to the 
integrand of Eqn.\,\ref{photon_Bbar} evaluated at $\left \{\Phi_{B},\Phi_{R}\right \}$. 
\end{enumerate}

%
%
\section{The generation of the hardest emission}\label{sec3}
Following the generation of the Born kinematics with next-to-leading-order accuracy
the hardest QCD or QED emission must be generated according to Eqns.\,\ref{eqn:qcd_sud}
or \ref{eqn:qed_sud}, respectively depending on whether a direct or photon fragmentation
contribution was selected.

\subsection{The hardest QED emission}

The hardest QED emission is generated by using the modified Sudakov form factor
defined in Eqn.~\ref{eqn:qed_sud}. We generate it in terms of the variables $\Phi_{R}^{\prime}(x_{p},z_{p},\phi)$, with
\begin{equation}
\mathrm{d}\Phi_{R}^{\prime} = \frac{1}{2\pi}\mathrm{d}x_{p}\mathrm{d}z_{p}\mathrm{d}\phi\text{,}
\end{equation}
defined in \cite{Seymour:1994ti, Seymour:1994we}, where $x_{p}\in[x_{\rm o},1]$, $z_{p}\in[0,1]$
and the azimuthal angle $\phi\in[0,2\pi]$. The invariant mass of the initial-final dipole $q^2=(p_i-p_k)^2=-Q^2$
is preserved by the photon radiation. It is easiest to generate the hardest emission
by introducing $x_\perp$ such that the transverse momentum of the emission relative to the
direction of the partons in the Breit frame of the dipole is $p_T=\frac{Q}2x_\perp$, where
\begin{equation}
x_{\perp}^{2}=\frac{4(1-x_{p})(1-z_{p})z_{p}}{x_{p}}\text{.}
\end{equation}
The Sudakov form factor can then be calculated in terms of $\tilde{\Phi}_{R}^{\prime}(x_{\perp},z_{p},\phi)$,
such that the $\theta$-function simply gives $x_{\perp}$ as integration limits and Eqn.~\ref{eqn:qed_sud} becomes
\begin{equation}
\Delta_{\rm QED}^{J}(x_{\perp})=\exp \left(-\int_{x_{\perp}}^{x_{\perp}^{\rm{max}}} 
\frac{{\rm d}x^\prime_\perp}{x_{\perp}^{\prime3}} {\rm d}\phi {\rm d}z_p
\frac{\alpha}{2\pi} \mathcal{W}\frac{\mathcal{A}_{\rm QED}^J}{B}\right)\text{,}
\label{eqn:qed_sud2}
\end{equation}
where
\begin{equation}
\frac{\alpha}{2\pi}\mathcal{A}_{\rm QED}^J=\frac{|D^{J}_{\rm{QED}}|}{\sum_{j}|D^{j}_{\rm{QED}}|+\sum_{j}|D_{\rm{QCD}}^{j}|}R(\Phi_{B},\Phi^J_{R})\text{,}
\end{equation}
the Jacobian, $\mathcal{W}$, is
\begin{equation}
\mathcal{W}=4z_p(1-z_p)(1-x_p)^2\text{,}
\end{equation}
and $\frac{Q}2x_{\perp}^{\rm{max}}$ is the maximum value for the transverse momentum. 
 
It is impossible to generate the hardest emission directly using Eqn.\,\ref{eqn:qed_sud2}
instead we use an overestimate
\begin{equation}
g(x_\perp)=\frac{a}{x_{\perp}^{3}}\text{,}
\end{equation}
of the integrand in Eqn.\,\ref{eqn:qed_sud2}
so that
\begin{equation}
\Delta^{\rm over}_{\rm QED}(x_{\perp})=\exp \left(-\int_{x_{\perp}}^{x_{\perp}^{\rm{max}}} 
\frac{{\rm d}x^\prime_\perp}{x^{\prime3}_\perp} {\rm d}\phi {\rm d}z_p 
a\right)
\end{equation}
can be easily integrated in $\left\{x_{\perp},x_{\perp}^{\rm{max}}\right\}$. This
allows us to solve $\mathcal{R}_1=\Delta^{\rm over}_{\rm QED}(x_{\perp})$ 
where $\mathcal{R}_{1}$ is a random number in $[0,1]$ to get the transverse momentum of
a trial hard emission
\begin{equation}
x_{\perp}^{2}(\mathcal{R}_1)=\frac{1}{\frac{1}{(x_{\perp}^{\rm max})^{2}}-\frac{2}{a}\ln{\mathcal{R}_{1}}}\text{.}
\end{equation}
This trial hard emission is then accepted or rejected using a probability given by the
ratio of the true integrand to the overestimated value. If the emission is rejected the procedure is
repeated with $x_{\perp}^{\rm max}$ set to the rejected $x_\perp$ value until the generated value is
below the cut-off. This procedure, called the \textit{veto algorithm}, correctly
generates the hardest emission according to Eqn.\,\ref{eqn:qed_sud2}~\cite{Sjostrand:2006za}.

%
%
\subsection{The hardest QCD emission}
The hardest QCD emission is generated in terms of the variables $\Phi_{R}(p_T,y,\phi)$ defined in Sect.~\ref{realemissionsect}. Eqn.~\ref{eqn:qcd_sud} then becomes
\begin{equation}
\Delta_{\rm QCD}^{I}(p_T)=\exp \left(-\int_{p_T}^{p_T^{\rm{max}}}
{\rm d}p^\prime_\perp {\rm d}\phi {\rm d}y
\frac{\mathcal{C}_{I}\alpha_s}{2\pi} \mathcal{W}_{I}\frac{\mathcal{A}^I_{\rm QCD}}{B}\right)\text{,}
\label{eqn:qcd_sud2}
\end{equation}
where 
\begin{equation}
\frac{\mathcal{C}_I\alpha_s}{2\pi}\mathcal{A}^I_{\rm QCD}=\frac{|D^{I}_{\rm{QCD}}|}{\sum_{j}|D^{j}_{\rm{QED}}|+\sum_{j}|D_{\rm{QCD}}^{j}|}R(\Phi_{B},\Phi^I_{R})\text{,}
\end{equation}
the Jacobian is
\begin{equation}
\mathcal{W}_{I}=\frac{x}{1-v_{\rm O}}\text{,}
\end{equation}
where we mean to use $v_{\oplus}$ for $I=\left\{qg; gq;g\bar{q}\right\}$ and $v_{\ominus}$ for $I=\left\{\bar{q}g\right\}$. 

As before we use the veto algorithm to generate the hardest QCD emission according to Eqn.\,\ref{eqn:qcd_sud2}.
In this case we introduce the overestimate function
\begin{equation}
g_{I}(p_T)=\frac{a_{I}}{p_T}\text{,}
\end{equation}
so that
\begin{equation}
\Delta^{\rm over}_{\rm QCD}(p_T)=\exp \left(-\int_{p_T}^{p_T^{\rm{max}}} 
\frac{{\rm d}p^{\prime}_T}{p^{\prime}_T} {\rm d}\phi {\rm d}y 
a_{I}\right)
\end{equation}
is easily integrable in $\left\{p_T,p_T^{\rm{max}}\right\}$ and $\mathcal{R}_1=\Delta^{\rm over}_{\rm QCD}(p_T)$ 
can be solved giving
\begin{equation}
p_T(\mathcal{R}_{1})=\mathcal{R}_{1}^{\frac1{a}}\text{.}
\end{equation}
As before this trial hard emission is then accepted or rejected using a probability given by the
ratio of the true integrand to the overestimated value. If the emission is rejected the procedure is
repeated with $p_T^{\rm max}$ set to the rejected $p_T$ value until the generated value is
below the cut-off. 

%
%
\section{Results}\label{sec4}

Unlike the implementations of many other processes in the POWHEG formalism it is impossible to 
directly compare our results for any quantities directly with next-to-leading-order simulations
in order to test the implementation due to the very different treatment of the photon fragmentation
contribution. Instead we compare a simple observable, the rapidity of the photons, with the
next-to-leading-order program \textsf{DIPHOX} \cite{Binoth:1999qq} as a sanity check of our
results not expecting exact agreement, although the PDFs and electroweak parameters
were chosen to give exact agreement for the leading order $q\bar q \to \gamma\gamma$ process.

\begin{figure}  
\centering  
\includegraphics[angle=90,width=0.7\textwidth]{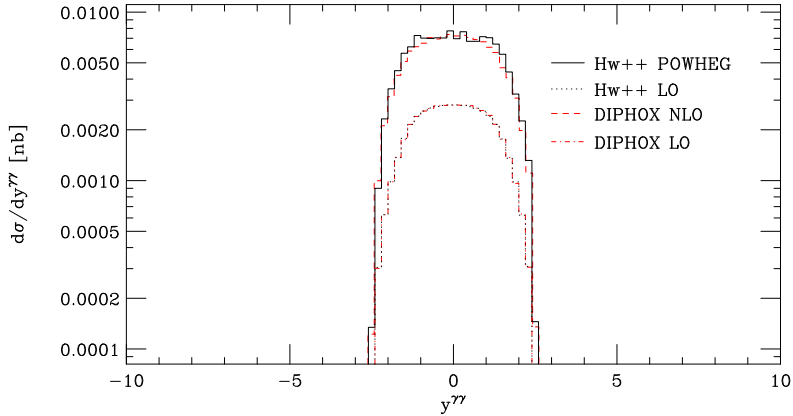}
\caption{Rapidity of the $\gamma\gamma$-pair at NLO. The distribution from the \textsf{Herwig++} parton shower with POWHEG correction (solid black line) is compared with NLO cross section from \textsf{DIPHOX} (dashed red line). At LO the \textsf{Herwig++} distribution is given by the dotted black line while the cross section from \textsf{DIPHOX} by the dash-dotted red line.}
\label{rapgamgam}
\end{figure}

For proton-proton collisions at a centre-of-mass energy of $14$ TeV, we used the following set of cuts on $p_T$ and rapidity of photons  
\begin{equation}
p_T^{\gamma}>25~{\rm GeV}, \hspace{2cm}|y^\gamma|<2.5,
\end{equation}
together with a cut on the invariant mass of the $\gamma\gamma$-pair  
\begin{equation}   
80~{\rm GeV}<M^{\gamma\gamma}<1500~{\rm GeV}\text{.}
\end{equation}
Moreover, we follow typical experimental selection cuts to isolate direct photons from the background:
we require that the amount of total transverse energy, $E_T^{\rm had}$, released in the cone,
centred around the photon direction in the rapidity and azimuthal angle plane, is smaller than $15$ GeV, \textit{i.e.}
\begin{eqnarray}
(y-y^{\gamma})^2+(\phi-\phi^{\gamma})^2&\le& R^2\\
E_T^{\rm had} &\le&~15~{\rm GeV}\text{,}
\end{eqnarray}
where $R=0.4$ is the radius of the cone. The PDFs are chosen to be the CTEQ6 set~\cite{Tung:2002vr}. 
The result is shown in Fig.~\ref{rapgamgam}. The distributions from \textsf{DIPHOX} at NLO(red dashed line) and LO (red dash-dotted line), together with LO \textsf{Herwig++} (dotted black line) and \textsf{Herwig++} with POWHEG corrections (solid black line) do not include the gluon-gluon channel. At LO the \textsf{Herwig++} and \textsf{DIPHOX} distributions are indistinguishable. At NLO they show a difference that is very small compared to the correction from LO to NLO, which means that the NLO curves are in reasonable agreement given the sizable contribution of the fragmentation contribution
that is treated differently in the two approaches.

\begin{figure}
\includegraphics[width=0.5\textwidth]{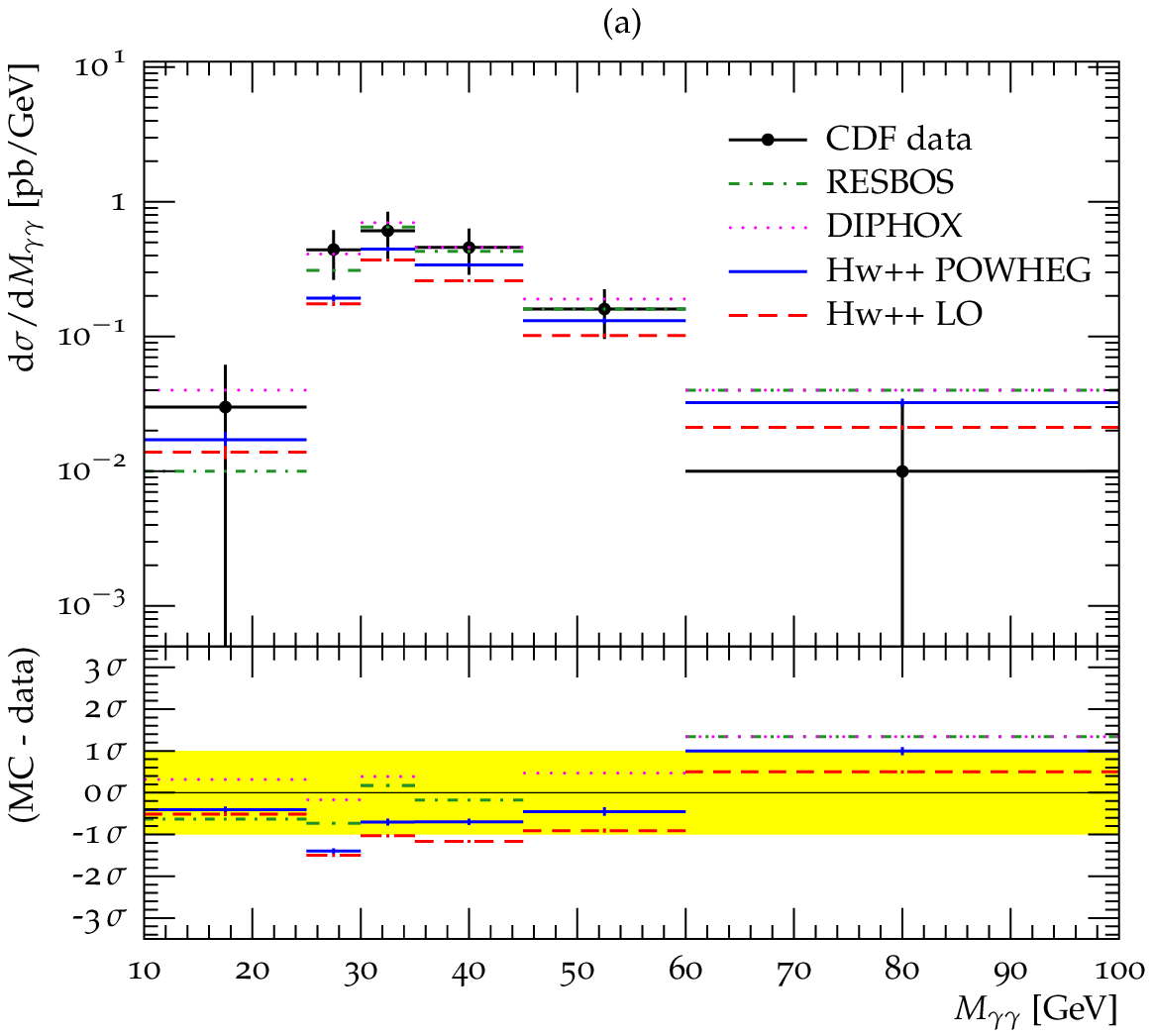}
\includegraphics[width=0.5\textwidth]{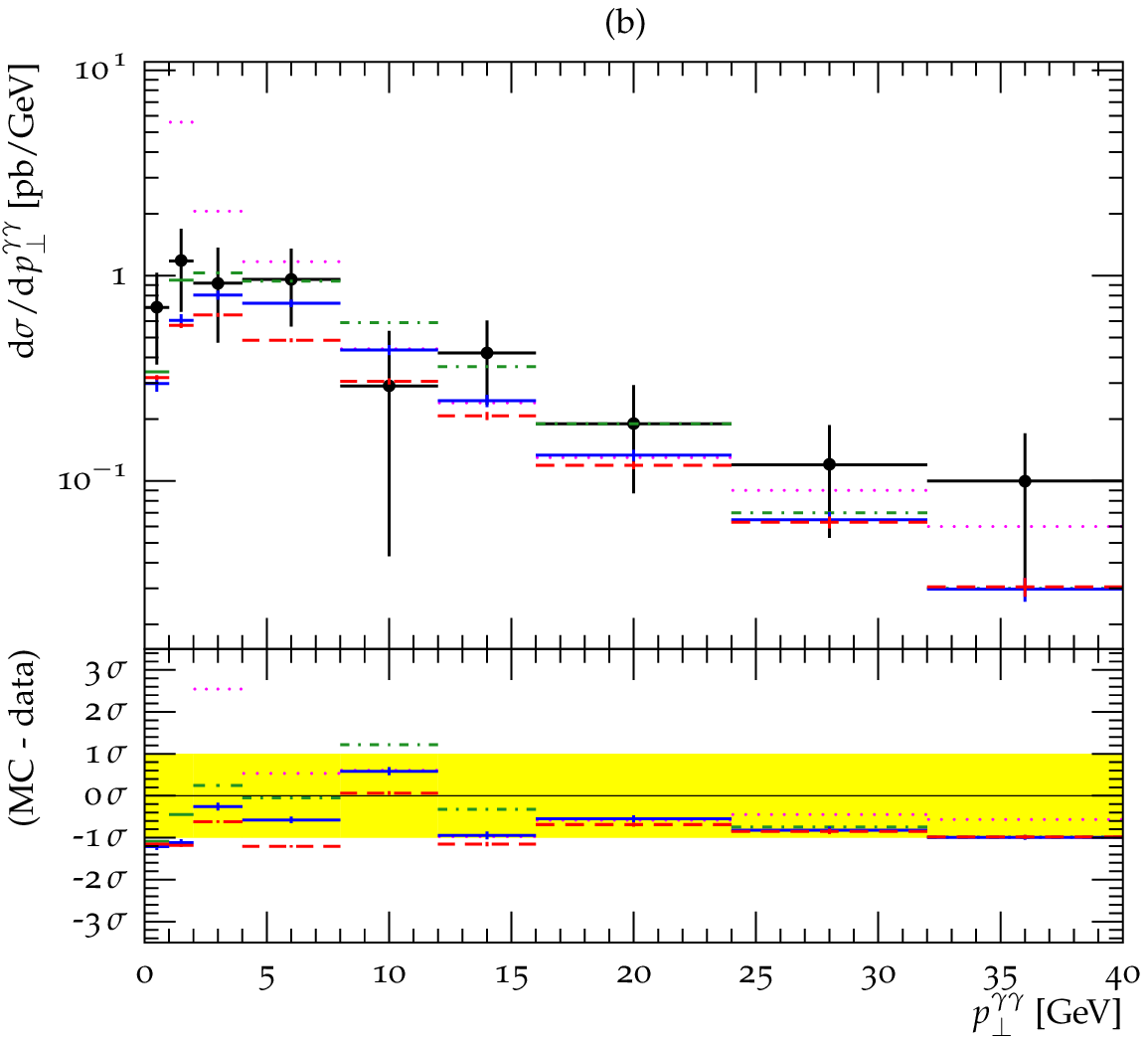}
\caption{The (a) invariant mass and (b) transverse momentum of the $\gamma\gamma$-pair.
         The solid blue line shows the POWHEG approach, while the dashed red curve shows the result of
         the \textsf{Herwig++} shower at LO. We show the NLO cross section provided by
         \textsf{DIPHOX} (magenta dotted line) and \textsf{RESBOS} (green dashed-dotted line).
         The data are from Ref.~\cite{Acosta:2004sn} and the curves are plotted with \textsf{Rivet} \cite{Buckley:2010ar}.
         In the lower panel, the yellow band describes the one sigma variation of data.}
\label{cdfplots1}  
\end{figure}

In Fig.\,\ref{cdfplots1}a we compare the results from \textsf{Herwig++} with the data
of Ref.~\cite{{Acosta:2004sn}}, a fixed next-to-leading-order calculation from \textsf{DIPHOX} (dotted magenta line)
and \textsf{RESBOS} (dashed-dotted green line) \cite{Balazs:1997hv,Balazs:2007hr,Nadolsky:2007ba,Nadolsky:2002gj,Balazs:1999yf}, which performs an analytic resummation of the logarithmically enhanced contributions.
Here and in the following the LO \textsf{Herwig++} parton shower (red dashed line) includes the $q\bar{q}\to\gamma\gamma$, $qg\to\gamma{\rm jet}$ and $gg\to\gamma\gamma$ contribution. The implementation of POWHEG correction improves the description and this results in a distribution (solid blue line) that is in good agreement
 with the data. Here, as in the following, the NLO curve includes the $gg\to\gamma\gamma$ subprocess. In the lower frame, we plot the ratio MC/data and the yellow band gives the
one sigma variation of data. All the plots comparing the results of \textsf{Herwig++} with
experimental results were made using the \textsf{Rivet} \cite{Buckley:2010ar} package.

It is of interest to study the transverse momentum of the $\gamma\gamma$-pair, because it is not infrared
 safe for $p_{\perp}^{\gamma\gamma}\to 0$. The $q\bar{q}\to\gamma\gamma$ and $gg\to\gamma\gamma$ processes
 present a loss of balance between the corresponding real emission and virtual contribution, which results
 in large logarithms at every order in perturbation theory. In addition, the fragmentation components
 introduce an extra convolution that smears out this singularity. Since \textsf{DIPHOX} is based on a
 fixed, finite order calculation it is not suitable for the study of infrared sensitive observables
 and it fails in the description of these observables at low $p_{\perp}^{\gamma\gamma}$, as it is shown
 in Fig.\,\ref{cdfplots1}b (dotted magenta line). Resummation for diphoton production in hadron-hadron
 collision has been provided at all orders in $\alpha_s$ in Ref.~\cite{Balazs:2006cc} and implemented
 in \textsf{RESBOS}, as the corresponding distribution (dashed-dotted green line) shows in the same
 figure. The \textsf{Herwig++} parton shower resums the effect of enhanced collinear emission to all
 orders in $\alpha_s$ in the leading-logarithmic~(LL) approximation and results in a finite behaviour for
 $p_{\perp}^{\gamma\gamma}\to 0$~(red dashed line).
 However, the LO distribution does not correctly describe the data. In presence of POWHEG
 correction the distribution (solid blue line) stays finite at low $p_{\perp}^{\gamma\gamma}$
 and is in good agreement with the CDF data~\cite{{Acosta:2004sn}}.    

\begin{figure}
\includegraphics[width=0.5\textwidth]{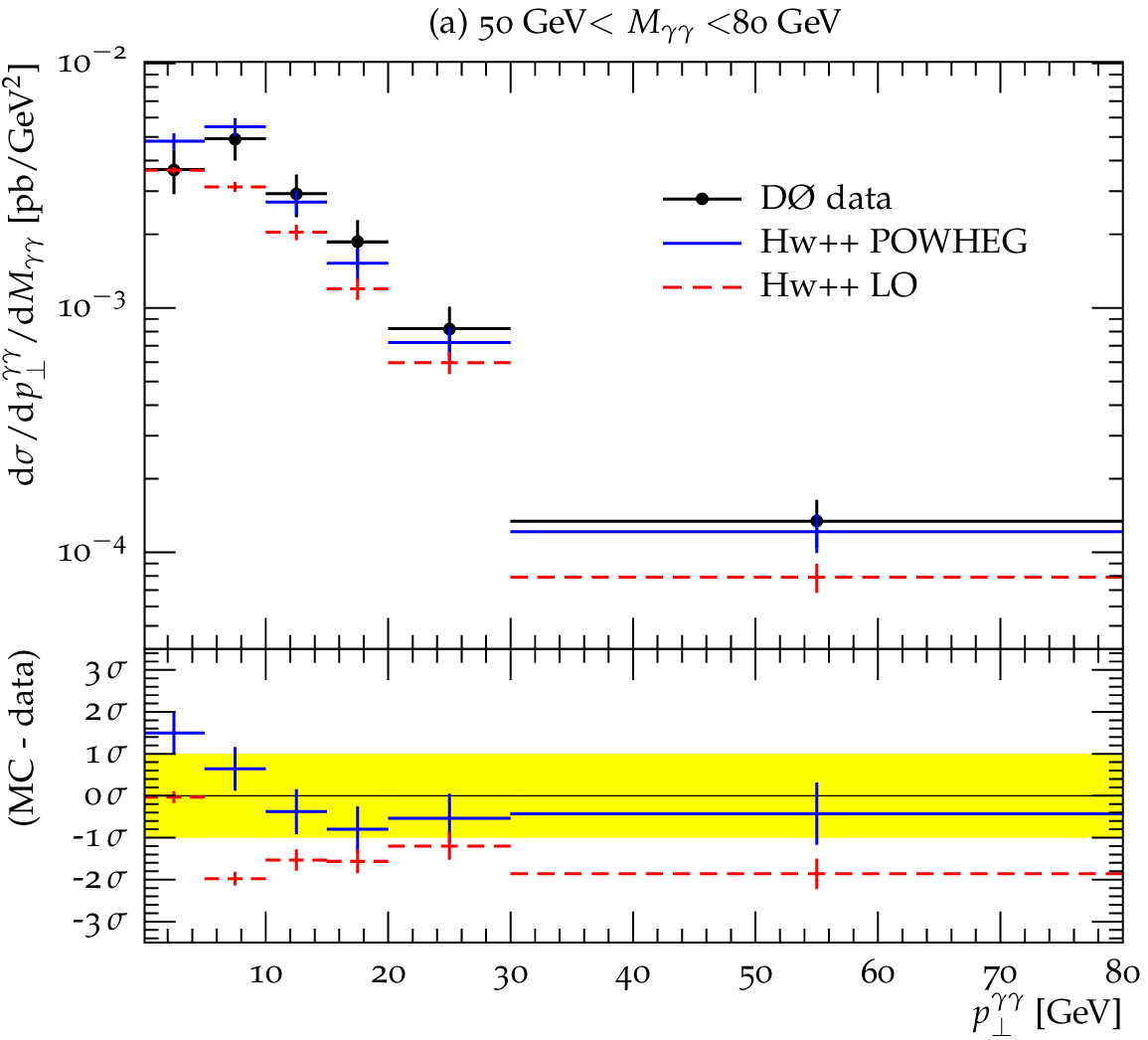}
\includegraphics[width=0.5\textwidth]{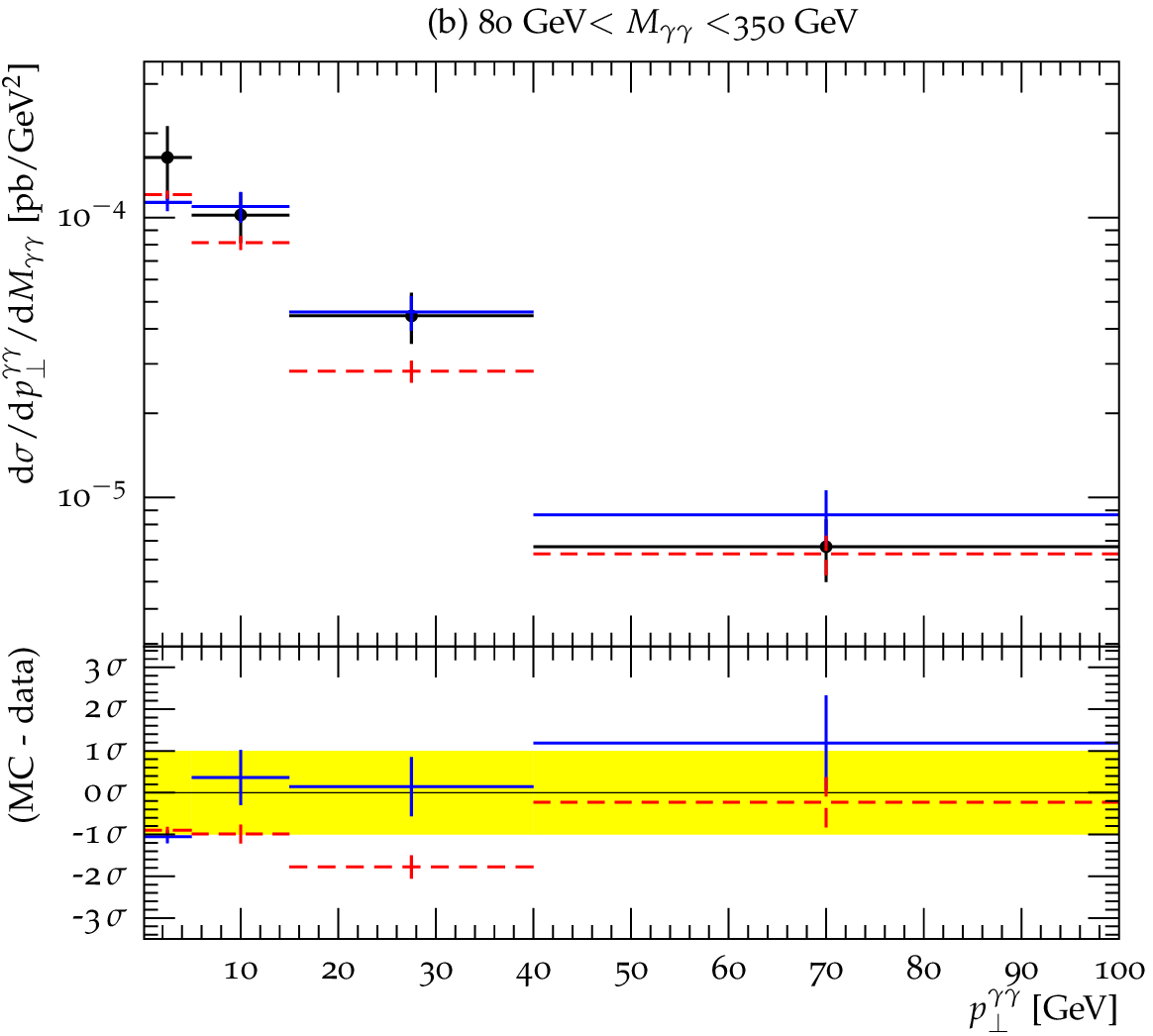}
\caption{Transverse momentum of the diphoton system for (a) $50{\rm~GeV}<M_{\gamma\gamma}<80{\rm~ GeV}$
         and (b) $80{\rm~GeV}<M_{\gamma\gamma}<350{\rm~ GeV}$. The distribution for the POWHEG formalism (solid blue line)
         is plotted together with the distribution for the \textsf{Herwig++} parton shower (dashed red line).
         The data are from Ref.~\cite{Acosta:2004sn} and the lower frame is as described in Fig.\,\ref{cdfplots1}} 
\label{d0plots1} 
\end{figure} 

\begin{figure}
\includegraphics[width=0.5\textwidth]{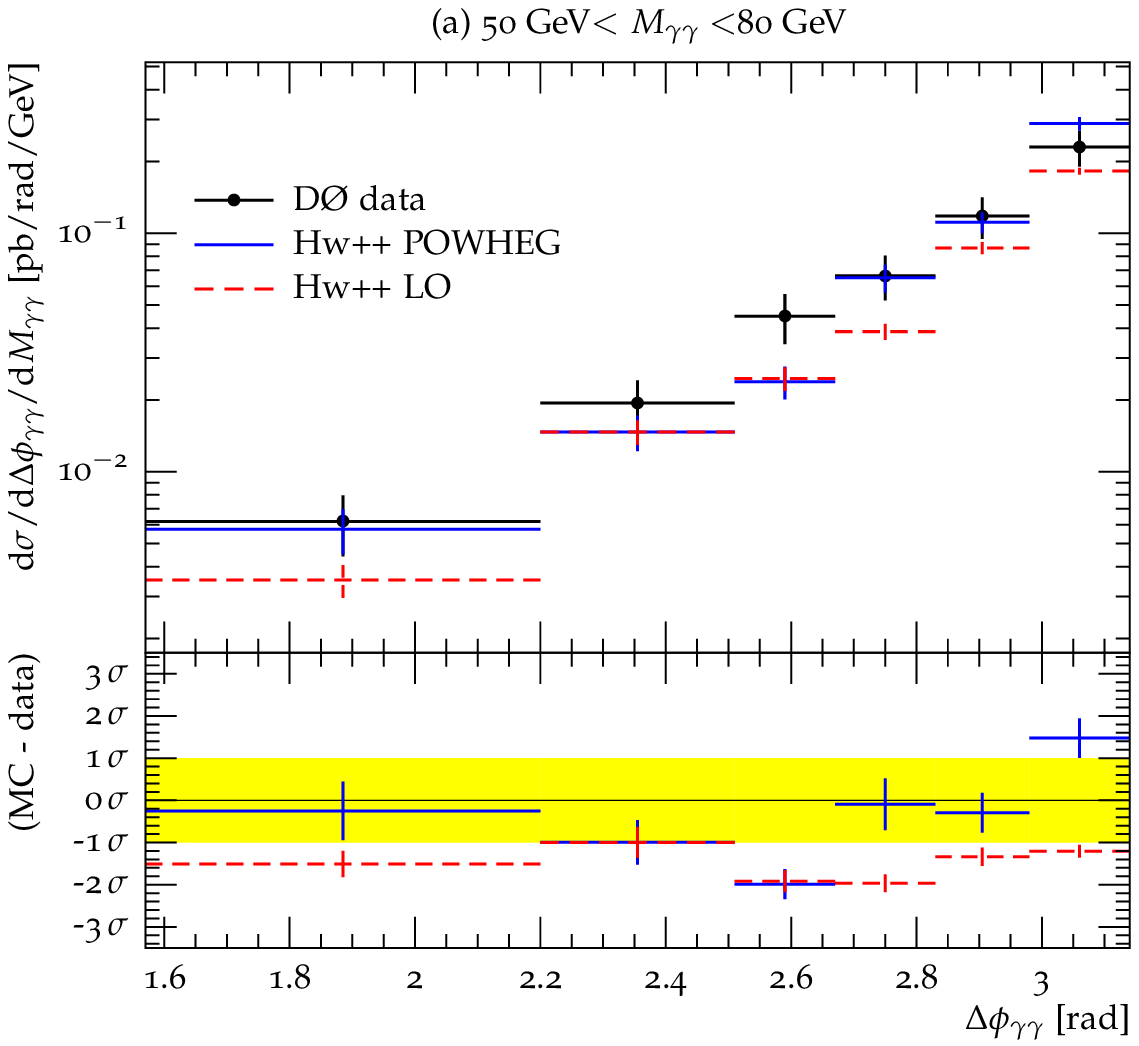}
\includegraphics[width=0.5\textwidth]{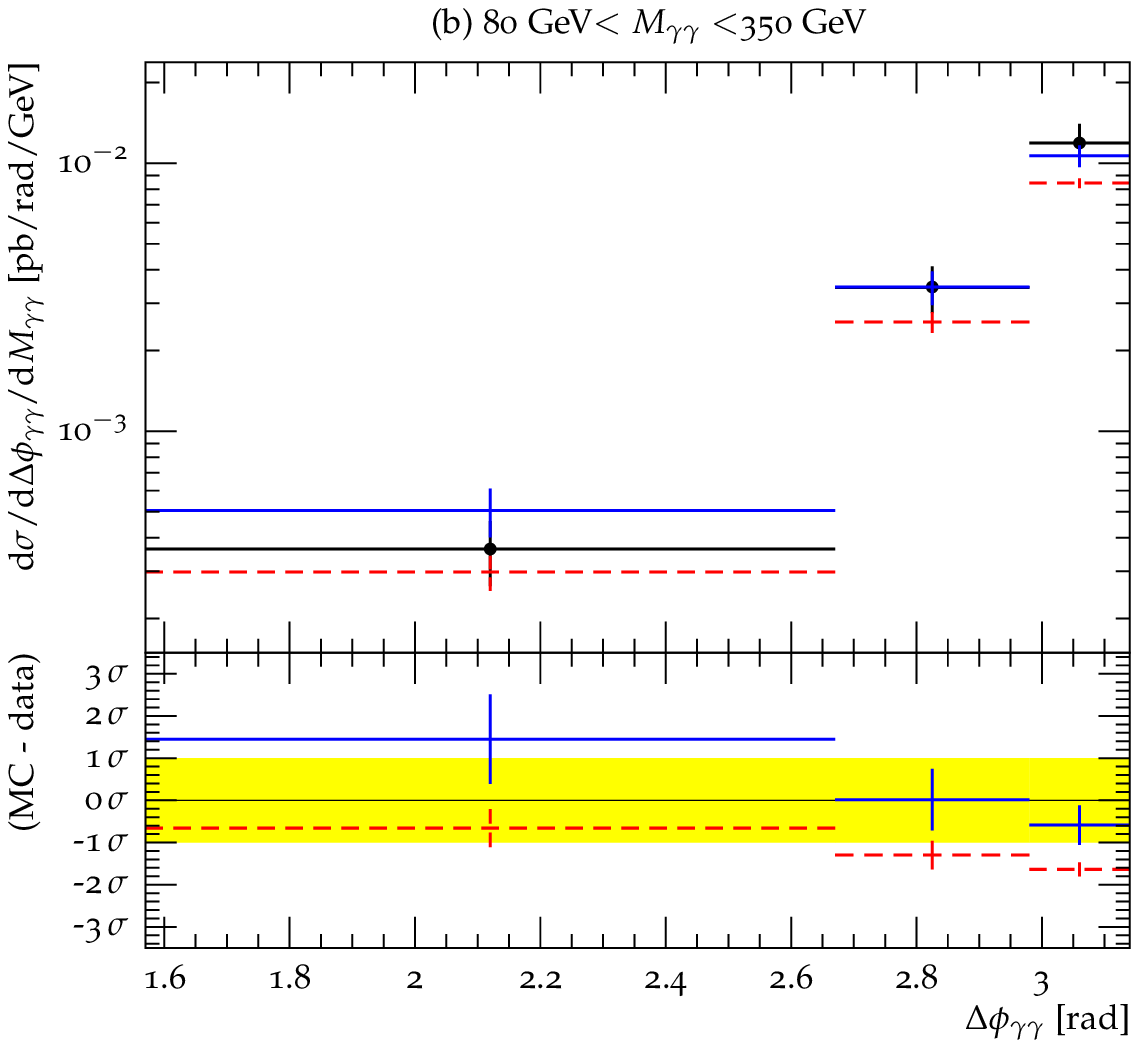}
\caption{Azimuthal angle between the photons for (a) $50{\rm~GeV}<M_{\gamma\gamma}<80{\rm~ GeV}$ and 
         (b) $80{\rm~GeV}<M_{\gamma\gamma}<350{\rm~ GeV}$. The solid blue line shows the result for the
         \textsf{Herwig++} shower with POWHEG corrections, while the red dashed line
         gives the result from the \textsf{Herwig++} parton shower.
         The data are from Ref.~\cite{Acosta:2004sn} and the lower frame is as described in Fig.\,\ref{cdfplots1}}
\label{d0plots2}
\end{figure}

\begin{figure}
\includegraphics[width=0.5\textwidth]{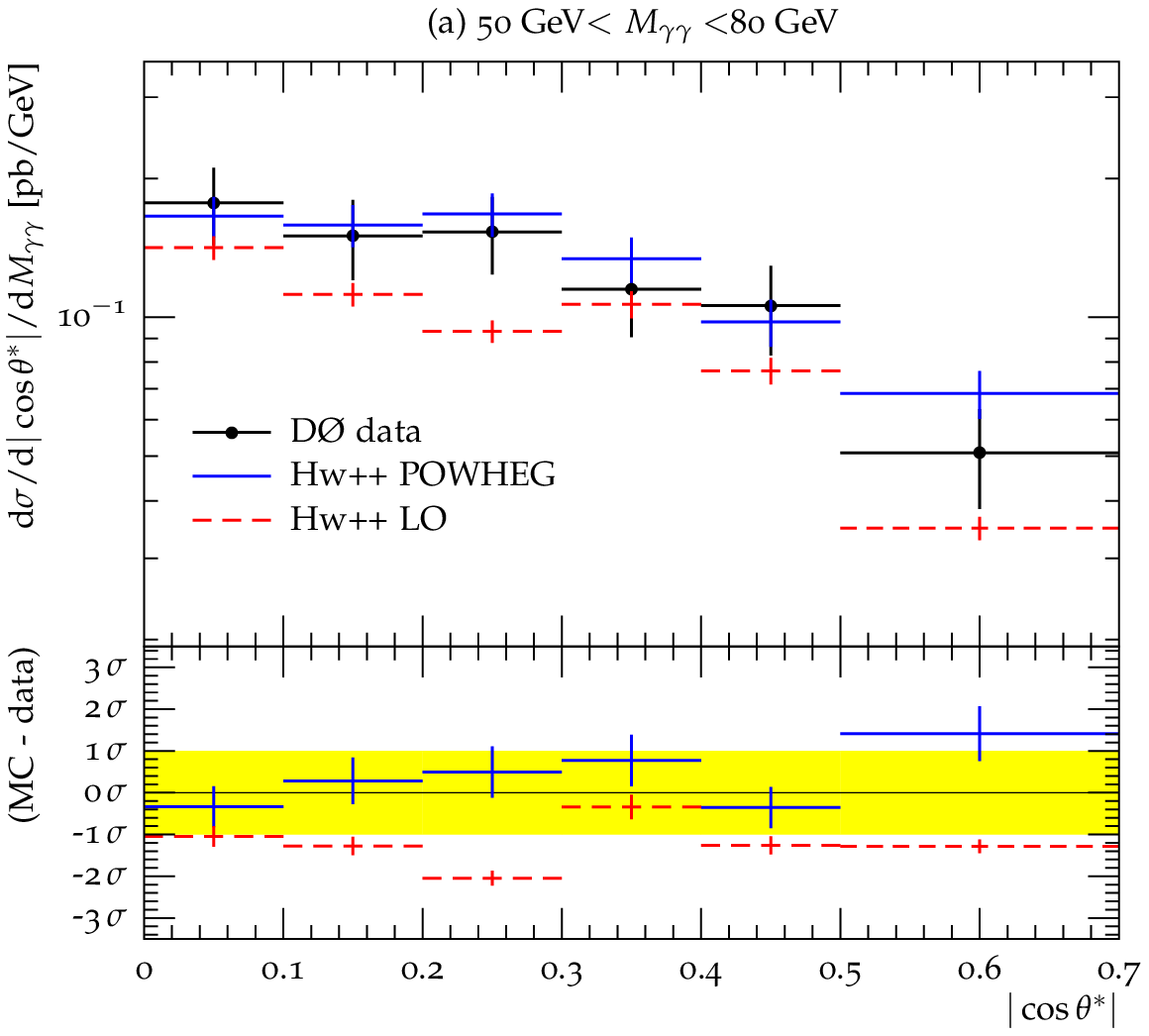}
\includegraphics[width=0.5\textwidth]{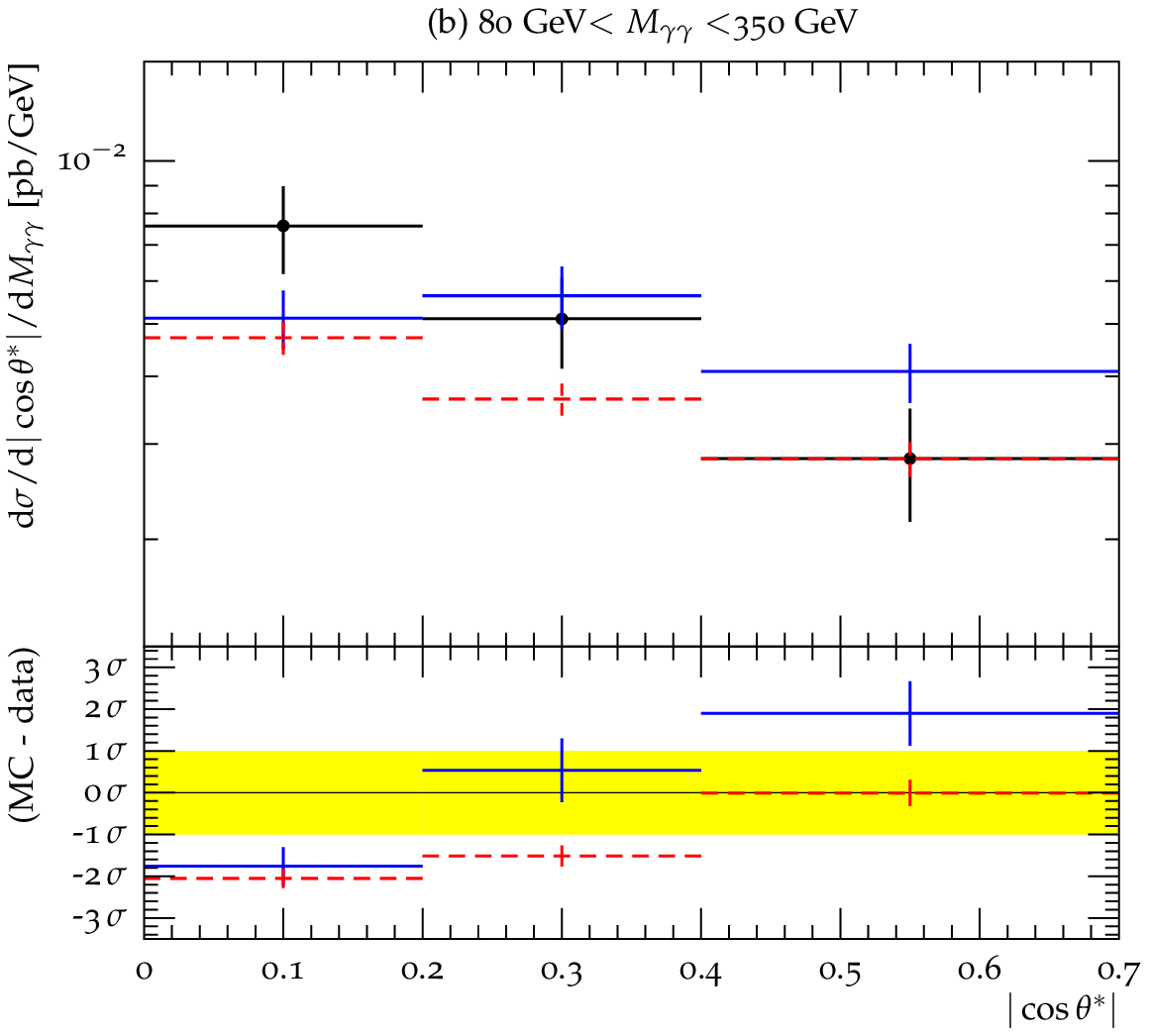}
\caption{Polar scattering angle between the photons for two ranges of $M_{\gamma\gamma}$: $50{\rm~GeV}<M_{\gamma\gamma}<80{\rm~ GeV}$ (a) and $80{\rm~GeV}<M_{\gamma\gamma}<350{\rm~ GeV}$ (b). The solid blue line describes the \textsf{Herwig++} result with POWHEG corrections, the dashed red line does not include matrix element corrections. The data are from Ref.~\cite{Acosta:2004sn} and the lower frame is as described in Fig.\,\ref{cdfplots1}.}
\label{d0plots3}
\end{figure}

In addition, \textsf{Herwig++} distributions, with and without POWHEG corrections,
are compared to the data of Ref.~\cite{Abazov:2010ah}. In Fig.\,\ref{d0plots1}, we
show the transverse momentum of the diphoton pair for two ranges of invariant mass
of the $\gamma\gamma$-pair, $M_{\gamma\gamma}$; in Fig.\,\ref{d0plots1}a $50{\rm~GeV}<M_{\gamma\gamma}<80{\rm~ GeV}$
and in Fig.\,\ref{d0plots1}b $80{\rm~GeV}<M_{\gamma\gamma}<350{\rm~ GeV}$. For the same ranges
of $M_{\gamma\gamma}$ we plot the azimuthal angle distribution between the photons in Fig.\,\ref{d0plots2}a
and Fig.\,\ref{d0plots2}b respectively and the polar angle between the photons in Fig.\,\ref{d0plots3}a
and Fig.\,\ref{d0plots3}b. For all distributions we see that the LO \textsf{Herwig++} ditributions (red dashed line) do not correctly describe the data. The POWHEG approach improves the simulation and provides a good description of D0 data~\cite{Abazov:2010ah}.

\section{Conclusion} \label{sec5}

In the present work the POWHEG NLO matching scheme has been extended and applied to $\gamma\gamma$-production
in hadron collisions. The QED singularities are not treated by including fragmentation functions but rather
by simulating the LO cross section for the corresponding process and then showering it. The simulation contains
a full treatment of the truncated shower which is needed to correctly generate radiation with transverse momentum
that is smaller than the one of the  hardest emission.

The implementation of the process was tested by comparing the results with the fixed-order \textsf{DIPHOX} program
which is in good agreement with the results of our approach for observables which are not sensitive to
multiple QCD radiation.

We find that without a correction to describe the hard QCD radiation there
is a deficit of radiation in the simulation. The POWHEG approach overcomes this problem and provides a
good description of the data of Refs.~\cite{Abazov:2010ah,Acosta:2004sn}. A remarkably good description is
obtained for infrared sensitive observables, like the transverse momentum of the $\gamma\gamma$-pair at
low $p_{\perp}^{\gamma\gamma}$, which demonstrates the resummation of logarithmic enhancement
provided by the \textsf{Herwig++} parton shower.  

This is the first NLO simulation of a process involving photons and provides an important new tool
for the study of promt photon production. The simulation will be made available in a forthcoming
version of the \textsf{Herwig++} simulation package.

\appendix
  
\bibliography{Herwig++}
\end{document}